\documentclass[runningheads]{llncs}
\usepackage{booktabs}
\usepackage{pgfplotstable}
\usepackage{adjustbox}
\usepackage{amsfonts}
\usepackage{amsmath}
\usepackage{amssymb}
\usepackage{bm}
\usepackage{float}
\usepackage{graphicx}
\usepackage{multirow}
\usepackage{subfig}
\usepackage{calc}
\usepackage{forloop}
\usepackage{tikz}
\usetikzlibrary{math}
\usetikzlibrary{fpu}
\usepackage{tablecoms}

\usepackage[colorlinks]{hyperref}

\graphicspath{{images/}}

\newcommand{\argmin}[1]{\underset{#1}{\operatorname{arg}\,\operatorname{min}}\;}

\begin{document}

\title{
Thermal Image Super-Resolution Using Second-Order Channel Attention with Varying Receptive Fields
}

\titlerunning{Second-Order Channel Attention with Varying Receptive Fields}                                                                                                        
\author{Nolan B. Gutierrez \and William J. Beksi
\thanks{
The authors acknowledge the Texas Advanced Computing Center (TACC) at the
University of Texas at Austin for providing software, computational, and
storage resources that have contributed to the research results reported within
this paper.}
}
\authorrunning{N.B. Gutierrez and W.J. Beksi}
\institute{University of Texas at Arlington, Arlington TX, USA\\
\email{nolan.gutierrez@mavs.uta.edu, william.beksi@uta.edu
}\\
}

\maketitle
\pagestyle{plain}                                                                                                                                             

\begin{abstract}
Thermal images model the long-infrared range of the electromagnetic spectrum and
provide meaningful information even when there is no visible illumination. Yet,
unlike imagery that represents radiation from the visible continuum, infrared
images are inherently low-resolution due to hardware constraints. The
restoration of thermal images is critical for applications that involve safety,
search and rescue, and military operations. In this paper, we introduce a system
to efficiently reconstruct thermal images. Specifically, we explore how to
effectively attend to contrasting receptive fields (RFs) where increasing the
RFs of a network can be computationally expensive. For this purpose, we
introduce a deep attention to varying receptive fields network (AVRFN). We
supply a gated convolutional layer with higher-order information extracted from
disparate RFs, whereby an RF is parameterized by a dilation rate. In this way,
the dilation rate can be tuned to use fewer parameters thus increasing the
efficacy of AVRFN. Our experimental results show an improvement over the state
of the art when compared against competing thermal image super-resolution (SR)
methods.
\end{abstract}


\section{Introduction}
\label{sec:introduction}
The purpose of single image super-resolution (SISR) restoration is to determine
the mapping between a possibly degraded low-resolution (LR) image and its
high-resolution (HR) counterpart. Finding this arrangement is difficult due to
the intractable nature of the problem. Techniques for discovering the mapping
can be divided into two areas: {\em interpolated} and {\em learning-based}.
Contemporary SISR has been dominated by deep learning which has demonstrated
superiority over hand-crafted methods such as bicubic and bilinear
interpolation.

Convolutional neural networks (CNNs) have been shown to be successful at
attending to visual images on tasks such as SISR. This includes
squeeze-and-excitation methods for global excitation of feature maps
\cite{hu2018squeeze,li2019compressing}, and the use of weight excitations
\cite{quader2020weight,lin2020context}. Furthermore, ablation studies have
conclusively shown that excitation networks for feature maps bring performance
gains \cite{zhang2018image}. On another note, a variety of methods exist for
modifying the RF of a CNN either through concatenation \cite{li2018multi},
deformation \cite{dai2017deformable}, or dilation \cite{szegedy2015going}. In
this work, we study how dilated convolutions offer \emph{compression through
dilations}. Concretely, we show how they modify the effective receptive fields
(ERFs) of a CNN and how they interact with an existing enhancement known as
second-order channel attention (SOCA) \cite{dai2019second}.
 
The development of infrared thermographic cameras has spurred researchers to
carry out innovative research in the thermal image domain. Representing
traditional SR, Mandanici et al. \cite{mandanici2019multi} combined geometric
registration with projection and interpolation to produce HR thermal images.
Naturally, researchers have recently investigated color-guided thermal image SR
\cite{chen2016color,almasri2018rgb}. For example, a pyramidal network provided
by Gupta et al. \cite{gupta2020pyramidal} attains accurate results by extracting
edge-maps from RGB images at various levels of the network.  Chudasama et al.
\cite{chudasama2020therisurnet} present an efficient SR network for thermal
images by eliciting high-frequency details with a limited number of feature
extraction modules. Other works have introduced popular loss functions to the
field of thermal image SR \cite{almasri2018multimodal,kansal2020multi}.

In this work we not only explore the use of efficient thermal SR, but we also
provide complete benchmarks on three thermal imagery datasets. Furthermore, we
compare the results of four architectural variants to assess performance gains
or losses due to the compression of parameters in our SR network. In contrast
to previous work on SISR, our proposed model applies SOCA to a concatenation of
features produced from convolutions with changing RFs. An intermediate
convolutional layer quantifies the importance of the values from each RF before
passing this information to SOCA. The key contributions of our work are the
following: (i) we show the effectiveness of SOCA for thermal image SR; (ii) we
present a novel approach to sample from a foveated RF; (iii) we demonstrate an
efficient network for multiple upscaling factors; (iv) we establish new
benchmarks on public thermal image datasets. Our source code is available at
\cite{attention_with_varying_receptive_fields_network}.

The rest of this paper is organized as follows. We provide a summary of related
thermal image SR work in Sec.~\ref{sec:related_work}, and the basics of SR in
Sec.~\ref{sec:preliminaries}. In
Sec.~\ref{sec:second-order_channel_attention_with_varying_receptive_fields}, we
propose an architecture for transforming LR input images to super-resolved
output images. In Sec.~\ref{sec:experimental_results}, we demonstrate our
dilation-rate driven deep attention to varying receptive fields network through
experimental results. Finally, we conclude in Sec.~\ref{sec:conclusion}.

\section{Related Work}
\label{sec:related_work}
Most recent developments in the image restoration domain focus on the visual
image space \cite{nasrollahi2014super,wang2020deep,anwar2020deep}. The known
importance of deeper CNNs in improving representational power has spurred the
development of architectures that improve stability and provide better
representations \cite{kim2016accurate,lim2017enhanced}. This is done not only
through more residual connections
\cite{kim2016deeply,anwar2020densely,liu2020residual}, but also through
structural preservation \cite{luo2016understanding,isobe2020video}, constrained
hypotheses spaces \cite{bahat2020explorable}, fast Fourier transform
\cite{reddy1996fft} and generative techniques \cite{saharia2021image}, and
student-teacher networks \cite{lee2020learning}. Additionally, SR works that
focus on improving contextual information have employed different attention
mechanisms \cite{mei2020image} and enhanced inception modules
\cite{muhammad2019multi}. 


\section{Preliminaries}
\label{sec:preliminaries}
The task of super-resolving an LR image to its HR counterpart may be summarized
by its image space and degradation model. Formally, the relationship can be
defined as 
\begin{equation}
  \bm{I}_x = D(\bm{I}_y;\bm{\theta}),
\end{equation}
where $D$ (known as the degradation function) maps an HR image $\textbf{I}_y$ to
an LR image $\textbf{I}_x$ with degradation parameters $\bm{\theta}$. Hence, SR
can be reduced to finding the parameters of $D$, however it is an intractable
process. Learning an SR model, $F$, can be formalized as 
\begin{equation}
  \bm{I}_y = F(\bm{I}_x;\bm{\theta}).
\end{equation}
Furthermore, any optimization algorithm can be applied to find these parameters
by minimizing an objective function,
\begin{equation}
  \hat{\bm{\theta}}= \argmin{\bm{\theta}} L(\hat{\bm{\theta}}_y,\bm{I}_y) + \lambda\psi(\bm{\theta}),
\end{equation}
where $\lambda$ is a small value that represents the importance of the
regularization term $\psi$. This term may aid a model's ability to generalize to
data never before seen. A common objective function to minimize is the mean
squared error (MSE) otherwise known as the 2-norm,
\begin{equation}
  \text{MSE}(\bm{I}_x,\bm{I}_y) =\frac{1}{N} \sum\limits_{i = 1}^{N}(\bm{I}_x(i) - \bm{I}_y(i))^2,
\end{equation}
where $N$ is the number of samples in a batch. 
 
\section{Second-Order Channel Attention with Varying Receptive Fields}
\label{sec:second-order_channel_attention_with_varying_receptive_fields}
\begin{figure}
\centering
\includegraphics[scale=.38]{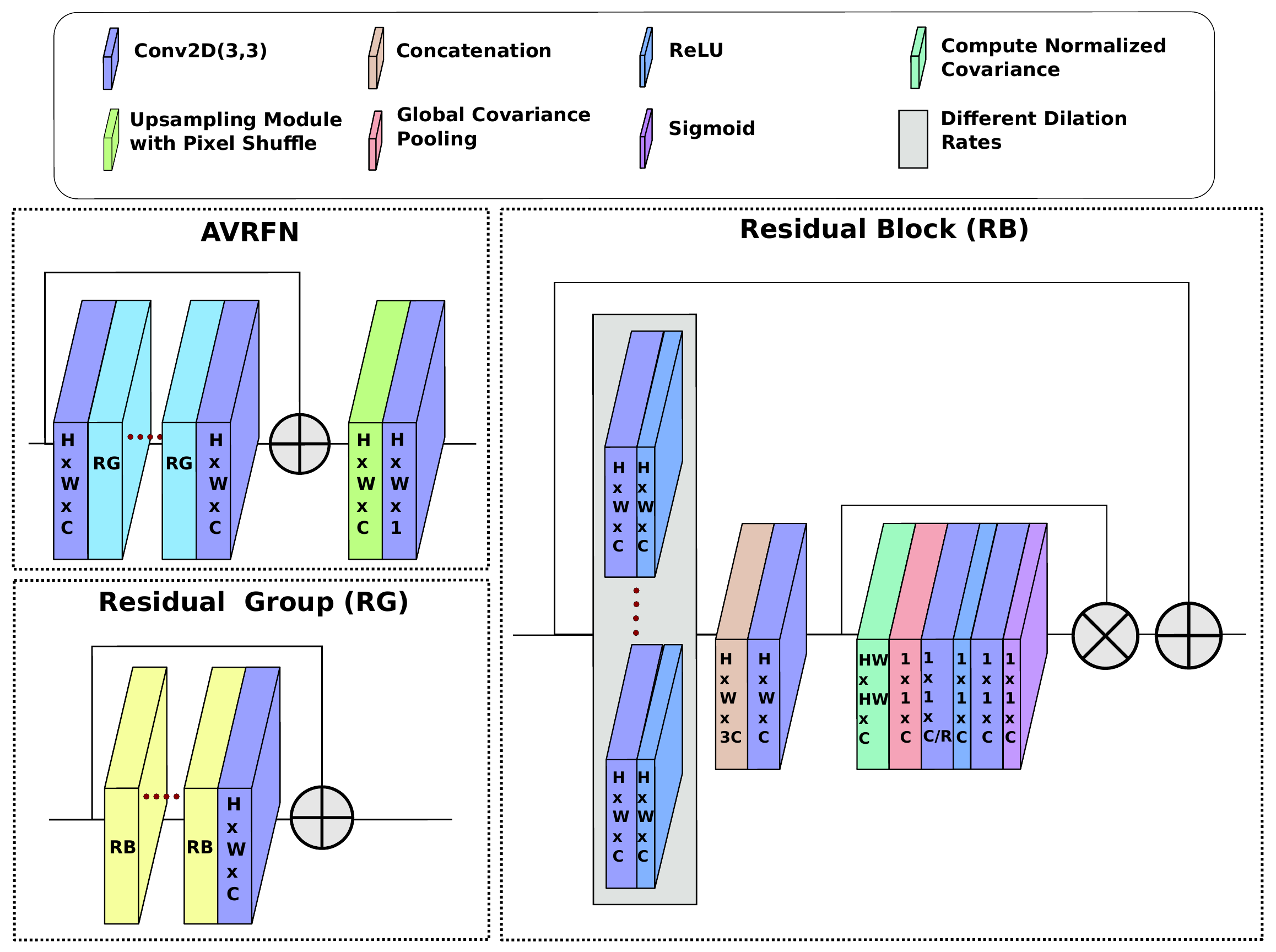}
\caption{An overview of our thermal imaging SR model.}
\label{fig:residual_body_2}
\end{figure}
We establish a novel deep learning architecture for thermal image SR as
follows. First (Sec.~\ref{subsec:second-order_channel_attention}), we provide a
detailed description of SOCA. Second (Sec.~\ref{subsec:dilated_convolutions}),
we expound upon how dilated convolutions are computed. Third
(Sec.~\ref{subsec:compression_through_dilations}), we describe how CNNs can be
compressed with respect to an ERF via compression through dilations. Fourth
(Sec.~\ref{subsec:residual_in_residual}), we incorporate residual in residual to
aid in the stabilization of our network. Finally
(Sec.~\ref{subsec:model_overview}), we apply SOCA to separate RFs (by dilation
rates) within residual in residual to assist in the SR of LR images obtained
with a bicubic degradation model. 

\subsection{Second-Order Channel Attention}
\label{subsec:second-order_channel_attention}
We utilize an alternative SOCA network to enhance convolutional blocks by
supplying a covariance matrix that allows for more discriminative
representations. To produce these second-order statistics, the covariance
normalization is obtained through Newton-Schulz iteration
\cite{higham2008functions}. Additionally, this serves to speed up the
computation. First, a feature map of dimension $H \times W \times C$ is
reshaped into a feature map $\bm{X}$ of shape $HW \times C$.  Second, the
covariance matrix is calculated, 
\begin{equation}
  \bm{\Sigma} = \bm{X}\bm{I_f}\bm{X}^\top,
\end{equation}
where $\bm{I_f} = \frac{1}{s}(\bm{I} - \frac{1}{s}\bm{1})$ and $s=HW$. $\bm{I}$ and 
$\bm{1}$ are the $m \times m$ identity matrix and the matrix of all ones, 
respectively. Then, the covariance matrix is pre-normalized, 
\begin{equation}
  \bm{\hat{\Sigma}} = \frac{1}{\text{tr}(\bm{\Sigma})}\bm{\Sigma},
\end{equation}
where tr($\cdot$) denotes the matrix trace. Let $\bm{Y}_0 = \bm{\hat{\Sigma}}$ 
and $\bm{Z}_0 = \bm{I},$ then $\bm{Y_n}$ and $\bm{Z}_{n}$ are obtained by
\begin{align}
 \bm{Y}_n &= \frac{1}{2}\bm{Y}_{n-1}(3\bm{I} - 
                  \bm{Z}_{n-1}\bm{Y}_{n-1}),\\
  \bm{Z}_n &= \frac{1}{2}(3\bm{I} -
                      \bm{Z}_{n-1}\bm{Y}_{n-1})\bm{Z}_{n-1},
\end{align}
with $\bm{Y_n}$ and $\bm{Z}_{n}$ quadratically converging to $\bm{Y}$ and
$\bm{Y}^{-1}$, respectively. The final normalized matrix after five iterations
of Newton-Schulz is found by compensating the pre-normalization step with
\begin{equation}
  \bm{\hat{Y}} = \sqrt{tr(\bm{\Sigma})}\bm{Y}_N.
\end{equation}
Afterwards, global covariance pooling is applied to obtain a scalar-valued
statistic $z_i$ for each channel $i$, 
\begin{equation}
  \bm{z}_i = \frac{1}{C}\sum\limits_j^C \bm{\hat{Y}}_{ij}.
\end{equation}
This permits the channel attention to capture correlations higher than the
first order. In the next step, the sigmoid activation function serves as a
gating mechanism that entrusts the network to selectively choose what to add to
the incoming input features. To create this gating mechanism, we use two
convolutional layers, $\bm{W}_0$ and $\bm{W}_1$, with rectified linear unit
(ReLU) and sigmoid activation functions. Concretely, 
\begin{equation}
  G(\bm{z}) = \bm{W}_1 \ast (\bm{W}_0 \ast \bm{z}),
\end{equation}
where $\ast$ is the convolution operation and $G(\bm{z})$ is an attention map.

\subsection{Dilated Convolutions}
\label{subsec:dilated_convolutions}
A dilated convolution multiplies a rate $l$ by $\Delta$ during the convolution 
operation, i.e.,
\begin{equation}
  (F\ast_lk)(\textbf{p}) = \sum\limits_{\Delta \in \Omega_r} F(\textbf{p} - l \cdot \Delta)k(\Delta),
  \label{eq:dilated_convolution}
\end{equation}
where $\Omega_r = [-r,r]^2 \cap \mathbb{Z}^2$, $k: \Omega_r \to \mathbb{R}$,
$\bm{p} \in \mathbb{Z}^2$ is a location on $\bm{X}$, $F: \mathbb{Z}^2 \to
\mathbb{R}$, and $*_l$ is an $l$-dilated convolution. In
\eqref{eq:dilated_convolution}, $k$ is known as the kernel function which
slides over $\bm{X}$. This allows a convolutional network to sample pixel
values from a larger RF over the input features. 

In the case of SR, it is advantageous to sample from different sized RFs 
depending on a number of factors including depth, resolution, and the 
scaling factor. We use dilated convolutions to extract features within each 
residual block. Specifically, the feature sets of three convolutional layers 
with varying dilation rates are concatenated and passed to an intermediate layer 
which pools the information from contrasting RFs. This intermediate layer 
effectively pools information at each feature map's location from a foveated RF 
where more parameters are concentrated towards the center of the field.

\subsection{Compression through Dilations}
\label{subsec:compression_through_dilations}
We utilize dilated convolutions to artificially increase the ERFs of our CNN. An
ERF is defined as the area containing any input pixel with a non-negligible
impact on a particular output unit within a feature map
\cite{luo2016understanding}. In addition, we introduce the concept of
compression through dilations as the case in which a CNN uses fewer parameters
to increase an ERF with dilated convolutions compared to without dilated
convolutions. For example, assuming we are using bias, two single-layer CNNs
defined by $\text{Conv2D}(\text{Input\_Shape}= (32,32,3),64,(5,5))$ and
$\text{Conv2D}(\text{Input\_Shape}=(32,32,3), 64,(3,3),\text{Dilation\_Rate}=2)$
have an ERF area of 25. However, our CNN has 160,064 and 57,664 parameters,
respectively, giving a compression ratio of 2.776.

\subsection{Residual in Residual}
\label{subsec:residual_in_residual}
We stabilize our deep channel attention network by the addition of residual in
residual (RIR) connections \cite{zhang2018image}. RIR entails two
levels of connections with groups on the outer level and blocks on the inner
level. More precisely, 
\begin{equation}
\hat{{\bf Y}} = {\bf X} + {\bf W} \ast R_g(R_{g-1}(\ldots R_1({\bf X})\ldots)).
\end{equation}
$R_g$ is the $g$-th residual group and it is formulated as
\begin{equation}
R_g({\bf X}) = {\bf X} + {\bf W} \ast B_t(B_{t -1}(\ldots B_1({\bf X})\ldots)),
\label{eq:residual_groups}
\end{equation}
where $B_t$ is the $t$-th residual channel attention block. We apply SOCA to
the features extracted from the convolutional layers with unique RFs. To obtain
the individual RFs, we make the dilation rate of each convolutional layer
exclusive and not equal to one for two of the layers. For example, if $\bf
W_1$, $\bf W_2$, and $\bf W_3$ are the weight sets associated with three
convolutional layers, then the residual block is derived as
\begin{equation}
B_t({\bf X}) = {\bf X} + \text{SOCA}([{\bf W_1} \ast {\bf X}, {\bf W_2} \ast_2 {\bf X}, {\bf W_3} \ast_3 {\bf X}])
\label{eq:residual_block}
\end{equation}
where $[{\bf W_1}, {\bf W_2}, {\bf W_3}]$ constitutes the concatenation of $\bf
W_1$, $\bf W_2$, and $\bf W_3$ along the channel axis.

\subsection{Model Overview}
\label{subsec:model_overview}
Our overall model is shown in Fig.~\ref{fig:residual_body_2}. To upscale an
input image, features are extracted from a series of residual groups and blocks
within the RIR architecture similar to RCAN \cite{zhang2018image}. Pixel
shuffle \cite{shi2016real} is used to rearrange $\bm{X}$ of shape $(H,W,C\cdot
r^2)$ to $(H\cdot r,W\cdot r,C)$ by periodically building a new feature map
$PS(\bm{X})$ with pixel values from dissimilar channels according to the
equation 
\begin{equation}
  PS(\bm{X})_{x,y,c} = \bm{X}_{\lfloor x / r \rfloor, \lfloor y / r \rfloor, C \cdot r \cdot mod(y,r) + C \cdot mod(x,r) + c}.
\end{equation}
Finally, a single-channel convolutional layer reduces the number of channels to
the same number as in the LR image.

\section{Experimental Results}
\label{sec:experimental_results}
Our experiments model the downscaling of HR thermal images using a bicubic
degradation model with statistical noise.

\subsection{Datasets}
We use the Thermal Image Super-Resolution (TISR) 2020 Challenge dataset, the
FLIR Thermal Dataset for Algorithm Training (TDAT) \cite{flir2021}, and the
KAIST multispectral pedestrian detection benchmark dataset
\cite{hwang2015multispectral}. The TISR dataset consists of three sets of 1,021
images from three distinct cameras. These cameras include a Domo, Axis, and
FLIR with a resolution of ($160 \times 120$), ($320 \times 240$), and ($640
\times 480$), respectively. Of these images, 60 were kept private, leaving 951
in the training set and 50 in the test set for each camera. For TDAT, we
evaluate on only the first 100 images captured by a FLIR FC-6320 camera.
Lastly, for the KAIST dataset, we collect every 200-th image from the day and
night scenes and then evaluate on the set of 226 images. The images from the
KAIST dataset were captured by a FLIR-A35 thermal camera with a resolution of
$640 \times 480$.

The ground-truth dataset was created by first forming batches of 16
single-channel image patches where each patch is of size $scale \times 48$. The
LR images were then obtained by bicubicly interpolating these patches to a size
of $48 \times 48$. For both training and testing, the images were preprocessed
by adding Gaussian noise with a mean of 0 and a variance of 10 for the bicubic
with noise degradation model. Finally, all elements of each LR patch are
normalized and clipped to [0,1].


\subsection{Implementation}
Our final architecture uses three residual groups with six residual blocks per
group. Each convolutional layer has 64 filters resulting in a highly-efficient
network. During training, the Adam optimizer \cite{kingma2014adam},
parameterized by a learning rate of $10^{-4}, \beta_1 = 0.9, \beta_2 = 0.999$,
and $\epsilon = 10^{-7}$, is applied to minimize the MSE of each batch for 300
epochs. Training with four NVIDIA GeForce GTX 1080 Ti GPUs took less than three 
hours per model. 


\subsection{Evaluation}
For the experiments, we tested four variants of our architecture to evaluate
the performance gains of the network. The variants are as follows.
\begin{itemize}
  \item Dilated residual in residual (DDRR): SOCA in the residual block of
  Fig.~\ref{fig:residual_body_2} is replaced by a convolutional layer with a $3
  \times 3$ kernel size and no activation function.
  \item Residual in residual with SOCA (RRSOCA): Our different dilation rate
  module in the residual block of Fig.~\ref{fig:residual_body_2} is replaced
  with a series of two convolutional layers each with a kernel size of $3
  \times 3$ and a ReLU activation function.
  \item Compressed RCAN (CRCAN) via dilated convolutions: This architecture is
  similar to RRSOCA, but the first and second convolutional layers have a
  dilation rate of 1 and 2, respectively.
  \item Attention to varying receptive fields network (AVRFN): This is our
  proposed model as shown in Fig.~\ref{fig:residual_body_2}.
\end{itemize}

\pgfplotstableread[col sep=comma]{data/thermal.csv}\data
\begin{table*}[ht]
 \centering
 \pgfplotstabletypeset[
 col sep=comma,
 columns={Test Set,Scale,Parameters,psnr,ssim},
 columns/Dataset/.style={column type={|c}, string type},
 columns/Test Set/.style={column type={|c}, string type},
 columns/Scale/.style={column type={|c}, string type},
 columns/Filters/.style={column type={|c}, string type},
 columns/Parameters/.style={column type={|c}, string type},
 columns/loss/.style={column type={|c}, string type},
 columns/psnr/.style={column name =PSNR,column type={|r},fixed zerofill, precision = 3},
 columns/ssim/.style={column name =SSIM,column type={|r|},fixed zerofill, precision = 3},
 columns/lossstd/.style={column type={|c}, string type},
 columns/psnrstd/.style={column name =$\sigma (\text{PSNR})$,column type={|r},fixed zerofill, precision = 3},
 columns/ssimstd/.style={column name =$\sigma (\text{SSIM})$,column type={|r|},fixed zerofill, precision = 3},
 every nth row={5}{before row = \bottomrule},
 every head row/.style={before row={\hline},after row=\hline},
 every last row/.style={after row=\hline},
 ]{\data}
\caption{The results of our AVRFN model on images captured by the TISR
\cite{rivadeneira2020thermal}, TDAT \cite{flir2021} and KAIST
\cite{hwang2015multispectral} datasets.}
\label{tab:thermal}
\end{table*}

\begin{figure}[ht]
\centering
\subfloat[]{
 \includegraphics[width=0.32\columnwidth]{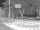}
 \label{fig:Input32_axis_4x}
}
\subfloat[24.43/0.64]{
\includegraphics[width=0.32\columnwidth]{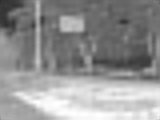}
\label{fig:Output32_axis_4x}
}
\subfloat[GT]{
\includegraphics[width=0.32\columnwidth]{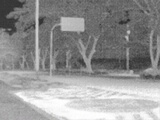}
\label{fig:10_domo_GT}
}\\
\subfloat[]{
 \includegraphics[width=0.32\columnwidth]{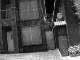}
 \label{fig:Input10_domo_4x}
}
\subfloat[26.64/0.83]{
\includegraphics[width=0.32\columnwidth]{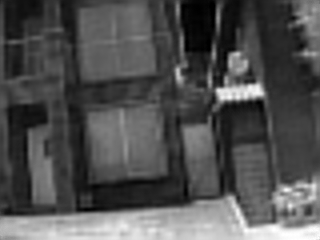}
\label{fig:Output10_domo_4x}
}
\subfloat[GT]{
\includegraphics[width=0.32\columnwidth]{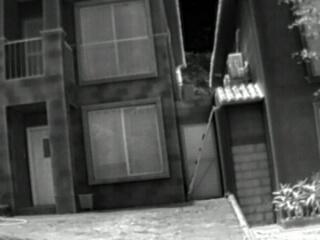}
\label{fig:32_axis_GT}
}\\
\subfloat[]{
 \includegraphics[width=0.32\columnwidth]{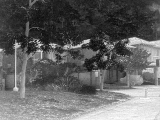}
 \label{fig:Input33_flir_4x}
}
\subfloat[28.54/0.78]{
\includegraphics[width=0.32\columnwidth]{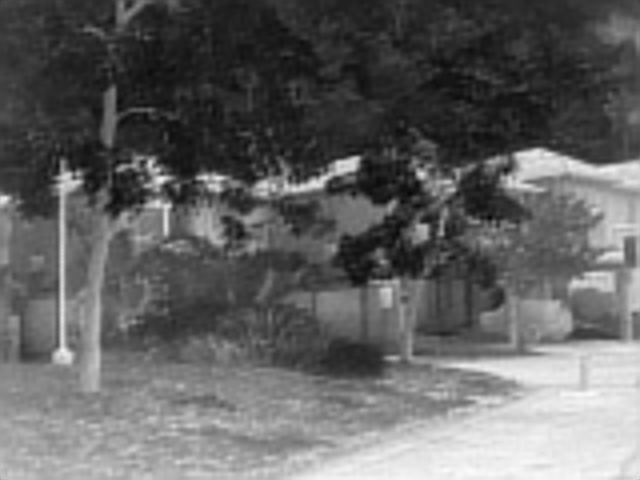}
\label{fig:Output33_flir_4x}
}
\subfloat[GT]{
\includegraphics[width=0.32\columnwidth]{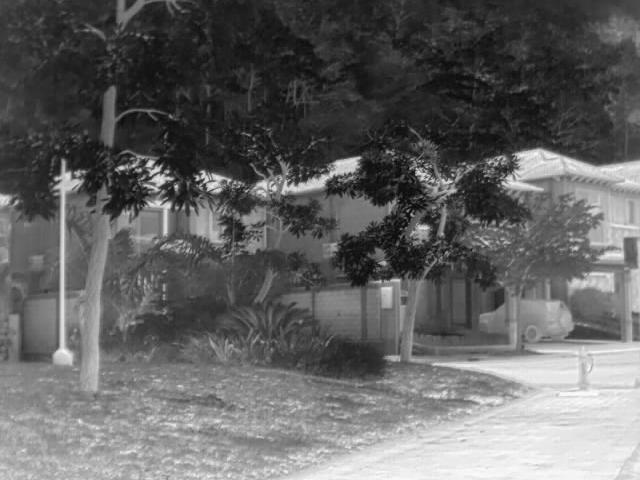}
\label{fig:33_flir_GT}
}
\caption{Examples of (left column) downsampled images from (top row)
low-resolution, (middle row) medium-resolution, and (bottom row)
high-resolution thermal cameras, their $\times 4$ upscaled counterparts (middle
column), and the ground truth (GT) (right column). Additionally, (b), (e), and
(h) show the PSNR and SSIM scores, respectively, when evaluated against the
ground truth.}
\label{fig:upscaling_4x_inputs_and_outputs}
\end{figure}

Fig.~\ref{fig:upscaling_4x_inputs_and_outputs} highlights example inputs, SR
predictions, and the ground-truth images from the datasets. The added Gaussian
noise produces heavily pixelated input images which presents very difficult
conditions to evaluate our methods on. In all of our experiments, we use the
peak signal-to-noise ratio (PSNR) and structural similarity index (SSIM) to
evaluate each architectural variant. Table~\ref{tab:thermal} shows the
performance of our proposed AVRFN when evaluated on each of the datasets.
Additionally, Table~\ref{tab:thermal_ablation} provides an ablation study which
compares the performance of the various types of compression with different
channel attention networks. Note that we were unable to make a fair comparison
with related work found in \cite{chudasama2020therisurnet} since the
performance of our baseline method did not match. However, each of our variants
performed better than the baseline RCAN architecture. 

An interesting finding is that adding compression through dilations in the
residual block of RCAN leads to improved performance. After each residual
connection, the ERF resets to the kernel size of the succeeding CNN due to the
easy pass-through of low-level information found in residual networks
\cite{araujo2019computing}. Contemporary work has found a correlation between
larger ERFs and performance gains \cite{araujo2019computing}. We hypothesize
that by introducing compression through dilations in each residual block, we
increase the ERF at a faster rate thus allowing for performance gains. An
unexpected result is that DDRR, the only variant without any form of
channel-attention, performs significantly worse. This confirms previous
ablation studies \cite{zhang2018image}, but it also shows that most of the
performance gains arise from channel attention and not compression through
dilations. In addition, our varying dilation rate module (AVRFN) improves
performance over the baseline which shows that attending to different RFs can
improve performance. Nonetheless, our highest performance gains are obtained
when we simply add compression through dilations to the RCAN baseline. 

\section{Conclusion}
\label{sec:conclusion}
In this work, we showed the advantage of attending to varying resolutions for
the reconstruction of thermal images by efficiently parametrizing a
convolutional layer with a dilation rate. Together with SOCA, our model achieves
state-of-the-art results on the task of thermal image SR and yields up-to-date
benchmarks for the research community.
In the future, we intend to look at ways in which training may be further 
stabilized and how attention to uncertainty maps can improve the computational 
efficiency of thermal image SR. 

\pgfplotstableread[col sep=comma]{data/thermal_ablation.csv}\dataone
\begin{table*}[ht]
 \centering
 \begin{filecontents*}{data/thermal_ablation.csv}
Dataset,Test Set,Scale,Filters,Parameters,loss,psnr,ssim,lossstd,psnrstd,ssimstd,model
thermal,\multirow{5}{*}{AXIS Domo P1290},4,64,1661377,219.38481,25.487497,0.6910673,131.27032,0.045636166,2.6760764,RRSOCA
thermal,,4,64,2839873,220.5252,25.458235,0.6914432,31.06572,0.04283732,2.666042,DDRR
thermal,,4,64,2839873,218.83322,25.490679,0.69220597,129.9528,0.043106087,2.6617856,CRCAN
thermal,,4,64,1661377,232.57706,25.23937,0.6824305,138.71713,0.046363495,2.6891055,RCAN
thermal,,4,64,1917313,225.48236,25.368172,0.68509763,133.87834,0.04725286,2.6828194,AVRFN
thermal,\multirow{5}{*}{AXIS Q2901},4,64,1661377,117.01289,28.166662,0.8024583,63.838524,0.05643561,2.61962,RRSOCA
thermal,,4,64,2839873,117.15253,28.158503,0.8012558,63.809887,0.05432983,2.6125484,DDRR
thermal,,4,64,2839873,116.31875,28.188532,0.8024526,63.087692,0.05424533,2.6134975,CRCAN
thermal,,4,64,1661377,123.330315,27.922888,0.79466194,66.70758,0.05761327,2.5867245,RCAN
thermal,,4,64,1917313,121.49283,27.989561,0.7969343,65.60109,0.057832636,2.5937788,AVRFN
thermal,\multirow{5}{*}{FLIR FC-6320},4,64,1661377,47.770023,31.977629,0.867,26.203661,0.0440749,2.368909, RRSOCA
thermal,,4,64,2839873,47.217308,31.985075,0.8669992,25.084635,0.040075492,2.282272,DDRR
thermal,,4,64,2839873,47.016727,32.001522,0.8672529,24.949545,0.040043693,2.2778952,CRCAN
thermal,,4,64,1661377,50.057632,31.755787,0.86066765,27.105618,0.044050172,2.3303084,RCAN
thermal,,4,64,1917313,49.51741,31.824211,0.863748,27.1874,0.045487214,2.375385,AVRFN
thermal,\multirow{5}{*}{TDAT},4,64,1661377,99.28229,28.387733,0.64080954,30.800293,0.049185194,1.4423307,RRSOCA
thermal,,4,64,2839873,98.24992,28.426987,0.645308,30.081879,0.047740392,1.4214623,DDRR
thermal,,4,64,2839873,98.28065,28.4256,0.64492327,30.08412,0.04778823,1.4215286,CRCAN
thermal,,4,64,1661377,101.962715,28.271002,0.63560164,31.660385,0.04890365,1.4367911,RCAN
thermal,,4,64,1917313,101.375206,28.298244,0.6365952,31.569386,0.049708117,1.445085,AVRFN
thermal,\multirow{5}{*}{KAIST},4,64,1661377,11.573302,37.976944,0.9487472,5.446628,0.018359492,2.0850985,RRSOCA
thermal,,4,64,2839873,12.725832,37.45552,0.91802084,5.3705587,0.03317716,1.8079995,DDRR
thermal,,4,64,2839873,12.4264,37.573235,0.92227006,5.3130245,0.030436117,1.8496269,CRCAN
thermal,,4,64,1661377,13.767239,37.088764,0.93751633,5.7536936,0.023043744,1.7183669,RCAN
thermal,,4,64,1917313,11.875478,37.82719,0.94262546,5.3944964,0.020322224,1.9950874,AVRFN
\end{filecontents*}
\pgfplotstabletypeset[
 col sep=comma,
 columns={Test Set,model,Scale,Parameters,psnr,ssim},
 columns/Dataset/.style={column type={|c}, string type},
 columns/Test Set/.style={column type={|c}, string type},
 columns/Scale/.style={column type={|c}, string type},
 columns/Filters/.style={column type={|c}, string type},
 columns/Parameters/.style={column type={|c}, string type},
 columns/loss/.style={column type={|c}, string type},
 columns/psnr/.style={column name =PSNR,column type={|r},fixed zerofill, precision = 3},
 columns/ssim/.style={column name =SSIM,column type={|r|},fixed zerofill, precision = 3},
 columns/lossstd/.style={column type={|c}, string type},
 columns/psnrstd/.style={column name =$\sigma (\text{PSNR})$,column type={|r},fixed zerofill, precision = 3},
 columns/ssimstd/.style={column name =$\sigma (\text{SSIM})$,column type={|r|},fixed zerofill, precision = 3},
 columns/model/.style={column name = Model, column type={|c}, string type},
 every nth row={5}{before row = \bottomrule},
 every head row/.style={before row={\hline},after row=\hline},
 every last row/.style={after row=\hline},
    highlight col max={\dataone}{ssim},
   highlight col max1={\dataone}{ssim},
   highlight col max2={\dataone}{ssim},
   highlight col max3={\dataone}{ssim},
   highlight col max4={\dataone}{ssim},
    highlight col max={\dataone}{psnr},
   highlight col max1={\dataone}{psnr},
   highlight col max2={\dataone}{psnr},
   highlight col max3={\dataone}{psnr},
   highlight col max4={\dataone}{psnr},
 ]{data/thermal_ablation.csv}
\caption{The results of our $\times 4$ model variants on images captured by
the TISR \cite{rivadeneira2020thermal}, TDAT \cite{flir2021} and
KAIST \cite{hwang2015multispectral} datasets.}
\label{tab:thermal_ablation}
\end{table*}

\clearpage
\newpage
\bibliographystyle{splncs04}
\bibliography{thermal_image_super-resolution_using_second-order_channel_attention_with_varying_receptive_fields}

\end{document}